\documentclass[aps,prb,preprint,groupedaddress]{revtex4}

\begin{document}
\title{The existence and properties of gases adsorbed within interstitial
channels of nanotube bundles}
\author{M. Mercedes Calbi}
\affiliation{Department of Physics, Southern Illinois University, Carbondale, Illinois 62901-4401}
\author{Flavio Toigo}
\affiliation{INFM and Dipartimento di Fisica "G. Galilei", via Marzolo 8, I-35131, Padova, Italy}
\author{Milton W. Cole}
\affiliation{Department of Physics, Pennsylvania State University, University Park, Pennsylvania 16802}
\date{\today}

\begin{abstract}

We analyze the relationship between our group's results and those of three recent 
papers about gases in interstitial channels within nanotube bundles. In particular, 
we explain the relation between our prediction about bundle dilation upon interstitial 
adsorption (questioned in some way in these works) and their results. 
\end{abstract}

\maketitle

During the decade since their first observation, carbon nanotubes
have provided a stimulus for extensive research, leading to many exciting
discoveries as well as a number of controversies. Recently, three studies have
explored different aspects of our group's work concerning gas adsorption
within and outside of nanotube bundles \cite{argon,boro,neon}. Each of these
papers draws conclusions that are, in some sense, critical of our group's
results \cite{dil,inter,rmp}. This Brief Report describes these studies very
briefly and explains the relation between our work and those studies. The
fact that these papers all appeared quite recently shows the importance of
the subject. Evidently, a mutual ``discussion'' of these controversial
questions is warranted. We believe that experimental data is necessary to
resolve some of these questions, but others can be answered without recourse
to experiments.

All three papers deal with the phases of adsorbed gases in interstitial
channels (ICs) within nanotube bundles. Numerous theoretical studies have
explored these phases for He, H$_{2}$ and Ne which are small enough to fit
comfortably within the IC \cite{boro,neon,dil,inter,rmp,uptake,siev,opt,boroqfs}. 
While several reports
concerning possible experimental observation of IC adsorption have appeared 
\cite{ye,bienfait,bob}, other articles have expressed contrary opinions
about their existence \cite{aldoic,aldoreview}. This difference may well be a
consequence of different kinds of experimental samples.

The three papers discussed here are related, in one way or another, to the
question of ``dilation'' of the nanotube bundle. The idea of dilation is
simply that a bundle of tubes will dilate, due to adsorption, in order to
accommodate the adsorbate. We evaluated the effect of dilation for the case
of H$_{2}$ in ICs and concluded that it can have a dramatic effect on the
ground state energy of the H$_{2}$ fluid\cite{dil,inter}.

One of the papers discussed here is by Gordillo, Boronat and Casulleras (GBC) 
\cite{boro}. This paper reports extensive, state-of-the-art investigations
of dilation of bundles containing H$_{2}$ at temperature $T=0$. As these
authors note, their results agree with ours for the case in which they
assume the same form of the C-H$_{2}$ pair interaction and the same force
constant of the inter-tube interaction \cite{ari,intertube}. One of the main
points made by GBC, however, is that the parameters of the C-H$_2$ interaction
might well differ from what was assumed in our work. In that case, they find
quite different results. What GBC omitted to mention is that we noted
exactly the same point in our original paper \cite{dil}, from which we quote:
``One factor to bear in mind is the sensitivity of these results to the
potential parameters. As examples, in the case of both H$_{2}$ and CH$_{4}$,
a 2.5\% decrease in the gas-carbon length parameter.... results in about a
25\% increase in the magnitude of [the chemical potential]''. In our second
paper on the subject \cite{inter}, we presented a graph showing this
dependence on the value of this uncertain parameter. The fact is that all
such theoretical studies rely on inadequately known interactions; the
dilation phenomenon is one for which the experiment is a exquisitely
sensitive probe of these interactions. Thus, there is no conceptual or
theoretical controversy but there does exist an open question that can be
addressed in future experimental studies. We note, in addition, that both
our work and GBC employ a simplified model in which the nanotube-molecule
interaction neglects the corrugation of the potential. While this neglect is usually 
well-justified in the case of adsorption on planar graphite (for which the
band structure correction to the effective mass is small \cite{graphcorr}), 
one study of He in the IC yielded a huge mass
correction \cite{vinic} ($m_{band}$ /$m_{bare}$ as high as a factor 20); such
a large corrugation drastically changes the energetics of adsorption.
Incorporating this effect requires knowledge of the chirality of the tubes
forming each IC, which is not known in the case of current experiments.
Until we have resolved these open questions, quantitative predictions are 
uncertain.

There is one further point made in the GBC paper upon which we comment.
These authors affirm that ``Probably, even if such effect [dilation] is
indeed present, its manifestation could be hardly observed due to the
present experimental uncertainties.'' We respectfully disagree, because
neutron or X-ray diffraction is capable of measuring the predicted dilation, 
an example of which is discussed below. In analyzing such
an experiment, it is necessary to take into account the effect of a
polydisperse distribution of nanotube sizes \cite{tubedist}. In addition,
finite bundle size effects necessitate more careful modeling of the dilation problem
than has been done thus far.

The second paper upon which we comment is by Brualla and Gordillo (BG) 
\cite{neon}. These authors compute the gas-liquid critical temperature ($T_{c}$)
of an anisotropic fluid consisting of Ne atoms in ICs that experience
interchannel interactions. In their paper BG describe our paper's results \cite{dil} 
as follows: ``There is even an estimation of a gas-liquid
transition critical temperature in the case of Ne, using a 3D modified
anisotropic Ising model to get $T_{c}$ . A comparison between the number
obtained in that approximation ($T_{c}=63.8$ K) and the one in the present
work [12 K] shows that in the neon case that Ising model is completely
inadequate, at least with the parameters given in that reference.''

This description of and comparison with our calculations is misleading in
two important respects. One is that our calculations take into account the
predicted dilation; their calculations ignore this effect. The key
significance of the predicted dilation is that it greatly increases the
binding of the Ne fluid so that $T_{c}$ is much higher than is found without
taking dilation into account. Hence, it is not appropriate to compare the
two sets of results for $T_{c}$. The second issue pertains to our use of the
Ising model, referred to above as ``seriously inadequate'', because of the
alleged disagreement in $T_{c}$. In a set of papers \cite{anisohe,aniso}, we
have found that this Ising model gives semiquantitative agreement with a
more realistic continuum method, perturbation theory (which is well
justified in the case of weak interactions between molecules in neighboring
ICs \cite{boro}). In summary, the factor of five disagreement between our 
$T_{c}$ values of those of BG are attributable to the important role of
dilation within our theory (absent from theirs). If the dilation were
omitted from our theory (or absent in the real system), the value of $T_{c}$
predicted by our theory, with the Ising model, would be 11.4 K for the
(5,5) nanotube bundle case, nearly coincident with the simulation value (12 K) 
found by BG.

We turn next to recent experimental results of Bienfait et al \cite{argon}.
These authors concluded from their data between 77 K and 96 K that Ar atoms in
ICs do not lead to dilation of the bundles. Although this finding was
interpreted as disagreeing with our predictions, this is not actually the
case. Our prediction was that Ar within ICs would have a chemical potential 
$\mu _{IC}=-1270$ K at $T=0$. This value is much lower than the value of $\mu$ 
calculated in the absence of dilation, indicating that dilation would
occur (at $T=0$) if the atoms are located within ICs. However, there is no
reason to expect such preferential IC adsorption because, for Ar atoms, 
the groove region of the
external surface of the bundle is energetically more attractive. The
same potential energy model used in the IC predicts a single atom binding
energy \cite{uptake} to be $E_{1external}=-1550$ K. At low $T$, however, the
groove atoms form a condensed state with much higher binding energy and
chemical potential. Taking into account just nearest neighbor interactions
within the groove yields an estimate $\mu _{external}=-1650$ K. This groove
phase is the lowest free energy (and chemical potential) state predicted for
the case of Ar interacting with nanotubes. At higher pressure and chemical
potential, of course, less strongly bound phases are expected to form; we
have carried out simulation studies of these on the external surface \cite{csargon}. 
The so-called three-stripe phase was found to appear at $\mu_{3stripe}=-1270$ K, 
while the external surface's monolayer forms at $\mu_{mono}=-1040$ K (at $T=30$ K). 
Thus, the IC adsorption and accompanying dilation is predicted to be 
energetically comparable with the three-stripe phase at very low $T$ but we do 
not have any prediction for the dilation's behavior at higher $T$. Our intuition 
suggests that the confinement within the ICs is inherently a low entropy 
situation (because of relatively high excitation energy). Hence a phase (like 
the monolayer) which is somewhat
less favored energetically than the dilated IC may well become more favored
at finite $T$. However, we reiterate that our calculations comparing the
energies of these phases are sensitive to uncertain assumptions, which
require testing by such experiments as these.

In our opinion, the strongest evidence for the existence of IC adsorption comes from 
the large isotope effect in the isosteric heat reported by Tate et al \cite{bienfait} for 
H$_2$ and D$_2$ at the lowest coverages. Our modeling of these problems has yielded a 
difference in isosteric heats for the ICs that is somewhat smaller than that found 
experimentally: 140 K when dilation is included vs. about 200 K found in the experiments. 
For the grooves, in contrast, our model yields a value near 50 K that is completely 
incompatible with the experimental data. While the individual results for the isosteric 
heats of H$_2$ and D$_2$ are sensitive to the model potential used, the isotope 
energy {\em shift} tends to be particularly sensitive to the geometry of the confining space 
(as expected for zero-point energy). 

We summarize the situation as follows. All theoretical studies of nanotube
adsorption have made numerous simplifying assumptions about the geometry and
interactions. Small differences between different models can yield quite
different results because of the sensitivity of the energy within the IC
environment. On the theoretical side, there is no doubt that dilation occurs
for an IC phase at low $T$ (except with a unique choice of interaction
length), but careful studies at finite $T$ and finite pressure are needed in
order to make comparison with experiments. Finally, we note that the
theoretical calculations described here rely on the assumption of perfect,
identical and parallel nanotubes. This idealization must be sacrificed in
order to make quantitative comparison with experiments, especially in those
situations where energy differences are small or where the phenomena are
particularly sensitive to the assumptions \cite{hete}.

\begin{acknowledgements}
We are grateful to Michael Bienfait, Oscar Vilches and Aldo Migone 
for extensive discussion of these problems. This research has been supported 
by the National Science Foundation.
\end{acknowledgements}

\end{document}